# Accessibility evaluation of websites using WCAG tools and Cambridge Simulator

**Shashank Kumar**

**BMS College of Engineering, Bangalore, India**

**JeevithaShree DV**

**Indian Institute of Science, Bangalore, India**

**Pradipta Biswas***

**Indian Institute of Science, Bangalore, India**

*** Correspondence:**
Corresponding Author
pradipta@iisc.ac.in

**Abstract**

There is plethora of tools available for automatic evaluation of web accessibility with respect to WCAG. This paper compares a set of WCAG tools and their results in terms of ease of comprehension and implementation by web developers. The paper highlights accessibility issues that cannot be captured only through conformance to WCAG tools and propose additional methods to evaluate accessibility through an Inclusive User Model. We initially selected ten WCAG tools from W3 website and used a set of these tools on the landing pages of BBC and WHO websites. We compared their outcome in terms of commonality, differences, amount of details and usability. Finally, we briefly introduced the Inclusive User Model and demonstrated how simulation of user interaction can capture usability and accessibility issues that are not detected through WCAG analysis. The paper concludes with a proposal on a Common User Profile format that can be used to compare and contrast accessibility systems and services, and to simulate and personalize interaction for users with different range of abilities.

## 1 Introduction

Web accessibility is one of the most important aspects of building a website. It is an inclusive practice of ensuring that there are no barriers that prevent interaction with, or access to websites on the World Wide Web, by people with physical and situational disabilities, and socio-economic restrictions on bandwidth and speed [WWW 2020]. Web developers should make sure that their website is accessible to people of varying abilities, or at least by those for whom the website is designed. Web developers usually use 'Authoring and Evaluation' tools to create web content. To evade confusions regarding accessibility assessment of different websites, World Wide Web Consortium (W3C) proposed a set of guidelines [W3C 2020] and formed the Web Accessibility Initiative (WAI). These guidelines are critical for developers who tend to design and develop websites, for example, a person with blindness requires screen reader technology while color contrast should be taken care of for persons with blurred vision. Similarly, videos should have sign language and should be close captioned for people with hearing impairment. Additionally, systems and services developed for elderly or disabled people often find useful applications for their able-bodied counterparts – examples include mobile amplification control, which was originally developed for people with hearing problem but helpful in noisy environment, audio cassette version of books originally developed for blind people, standard of subtitling in television for deaf users and so on. However existing design practices often isolate elderly or disabled users, by considering them as users with special needs and do not consider their problems during the design phase. Later they tried to solve the problem by providing few accessibility features. In the context of accessibility of websites, this paper made the following contributions:

- Compared 10 different web accessibility tools on two popular websites
- Identified issues with existing WCAG tools in terms of adaptability, distinguishability, compatibility, navigability, formats and so on
- Used a simulator [Biswas et al. 2012] that can identify interaction issues for people with different range of abilities and complements results obtained from existing tools
- Proposed a common user profile format to personalize content of websites for people with different range of abilities.

## 2 Literature Review

The World Wide Web (WWW) was invented by Tim Berners Lee in 1989 [WWW 2020] to serve researchers with data across the globe. Later, in 1991, it was released for public use. Since then, the WWW has undergone several improvements to make it available to a larger population across the globe. However, there still exists problems of accessibility, which limits a major fraction of population from utilizing it. Accessibility evaluation of a website is a complicated and difficult task. Hence, most web developers fail to evaluate websites for accessibility in the website development life cycle, making it inaccessible for people with disabilities.

In early days, websites were evaluated manually, as tools were not reliable in flagging issues. An inspiring paper by Brajnik et al. [2011], discussed evaluation of accessibility checks done by experts resulting in detection of such violations with more accuracy. Manual inspection method consumed more manpower and time, thereby being expensive compared to automated tools. There is no way to confirm the expertise of an evaluator resulting in biased evaluations, and experts may fail to perform a detailed meticulous review. Another approach was to evaluate websites for accessibility by conducting user trials with people with disabilities. This method is known as user testing, where end users themselves provide an accurate assessment [Petrie and Kheir 2007]. However, this method is not easy, as it requires the need to find users with disabilities, every time a website needs to be evaluated.

Disabilities are of various types, so we would have to find appropriate user group, to ensure thorough evaluation. Nowadays, most automated web-based tools help in evaluating websites by providing a detailed report of areas which need to be improved [Kumar and Owston 2016].

In a recent study by Alsaeedi [2020], only two tools were used to check for accessibility of a website. Tools were selected without proper justification for their selection. A study by Sandhya and Devi [2011], focused on evaluating The Times of India and NDTV webpages, using a screen reader. The study just provided a discussion on methods followed in using screen reader and not the evaluation process itself. Abdullah [Alsaeedi 2020] evaluated the website of the University of Saudi Arabia for accessibility using two tools. Authors proposed a Coverage Error Ratio (CER) metric to select any web accessibility tool. It is the ratio of number of errors detected by a given tool to the total number of errors detected by all tools. Value of CER scores calculated for each tool, can be used to measure the performance of tools. In this study, authors used the Wave and Site Improve tool for webpage evaluation as they possessed a high CER score. In a recent study by Rafael and Carlos [2020], a webpage was evaluated by three tools namely Color Contrast Accessibility Validator, WAVE and WCAG Color Contrast Validator, to check if the webpage meets color contrast specifications. Tools used were compared based on different sets of foreground and background colors [Rafael and Carlos 2020].

In this paper, we have selected 10 tools, based on many crucial factors that affect results of website evaluation. Tools have been compared among themselves, in order to aid and expedite the work of future web evaluators and researchers. Additionally, we have used a simulation-based approach to further evaluate two websites for accessibility. The simulator is designed as a tool, to help web designers visualize, understand and measure effects of age and impairment, on their websites. Finally, we propose a common user profile format to personalize websites across different platforms and devices.

## 3 Evaluation of Websites Using Web Accessibility Tools

For our study, we have considered evaluation of landing pages of two popular websites- the 'World Health Organization (WHO)' and the 'British Broadcasting Corporation (BBC)'. The WHO is a part of the United Nations Organization, which is tasked with promoting universal healthcare, monitoring public health risks, coordinating responses to health crises, and bettering human health and wellbeing [WHO Website 2020]. The organization is headquartered at Geneva, Switzerland, with 6 semi-autonomous regional offices and 150 field offices worldwide [WHO Website 2020]. The BBC is the world's oldest media broadcaster headquartered in Westminster, London [BBC Website 2020]. It has a staff of around 22,000, and more than 16,000 other personnel who are engaged in this public sector broadcasting organization [BBC Website 2020]. The website hosts its online services in Arabic and Persian languages. In our study, we considered the following 10 web accessibility tools available on the W3C website for evaluation. Tools were selected based on features, supported formats and browsers.

- A-Tester by Evaluera Ltd
- A-Checker by Inclusive Design Research Centre
- Functional Accessibility Evaluator 2.0 by University of Illinois at Urbana-Champaign
- Contrast checker by Art
- WAVE by WebAIM
- Accessibility Insights for Web by Microsoft
- Button Contrast Checker by Aditus
- Siteimprove Accessibility Checker by Siteimprove
- Utilitilia Validator by Utilitia SP. Z O.O
- Who Can Use by Corey Ginnivan

## 4 Comparative Analysis of Web Accessibility Tools

Accessibility, usability, and inclusion are three important factors to be considered while developing a website usable by people with disabilities. The 'Web Accessibility Initiative' (WAI) suggests a set of guidelines to be followed by developers of any website. In our study, we considered 10 tools available on the W3C for evaluation process [W3C Tool List 2020]. Guidelines in the W3C are categorized into three levels of conformance in order to meet the needs of people of different groups and different situations. These levels are as follows:

- *Level A:* The website should have utilized meaningful sequences, color, and audio control. This is the most basic level that should be met, to get a positive accessibility result.
- *Level AA:* The website should meet all conditions mentioned in Level A. It should hold proper labels/headings, and minimum contrast should be present, along with audio descriptions.
- *Level AAA:* The website should successfully meet all the criteria mentioned in Level A & Level AA, extended audio descriptions, alternate media files, and interruptions.

Guidelines and success criteria defined by the W3C are organized around four principles. These principles lay the foundation necessary for anyone to access and use the Web content with ease [W3C Intro. 2020]. The Webs' content should be:

- *Perceivable:* Content present on website should be properly presented. User must be able to apprehend a maximum understanding of contents of website.
- *Operable:* All navigation and interface links/components must work properly. There should be no hassles in using website (no dummy button/links should be present).
- *Understandable*: User must understand offerings of website. The websites' content should be written in an easy to understand manner, or at least in a manner understandable by the targeted audience.
- *Robust:* Contents of website should work across multiple platforms, without any interface issues, and should provide same experience to all its users, across all platforms. Since the Web industry is improving on a day to day basis, websites should be kept updated to meet latest standards.

If any of these principles are not met, users with disabilities will find it difficult to use the webpage.

Our study is divided into two phases; *Phase – 1* describes 10 tools selected for accessibility evaluation of two websites. For each website, we have described results in details, issues identified and alternate suggestions to solve the issues. Selection of tools have been performed based on parameters described in section 4.1.2. *Phase – 2* evaluates the two websites further, using the Cambridge Simulator [Biswas et al. 2012]. This tool provides an overview of how the website would be accessible to a user with disability with the help of simulation. Website checks for three types of disabilities- visual, motor and hearing impairments. Results obtained from the simulator, for each disability, has been mentioned in section 4.2.1. A detailed description of the two phases are described below in sections 4.1 and 4.2 respectively.

### 4.1 Selection of Appropriate Tools to Evaluate Accessibility

This section presents results for evaluation of two websites using 10 tools available on the W3C. A detailed description of the 10 tools, along with results, are furnished below:

- ***A Tester by Evaluera Ltd*** [Release date: 28-May-2014]
  This tool checks webpages for WCAG 2.0 Level AA conformance guidelines in the HTML code.

| **Results** | **WHO** | **BBC** |
|---|---|---|
| **Error Chart** | 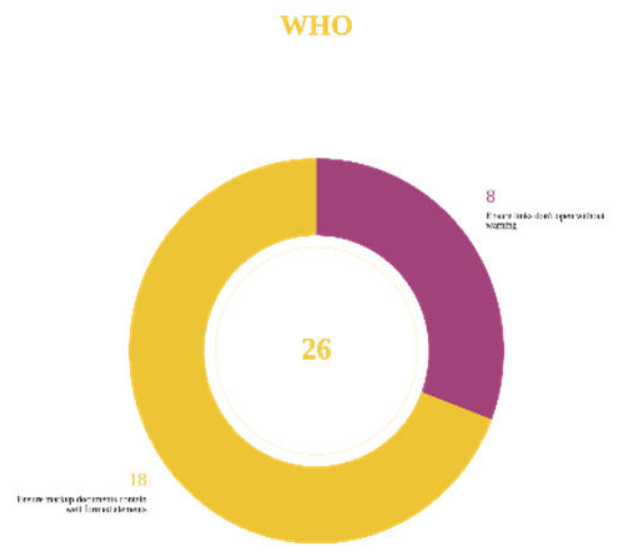 | 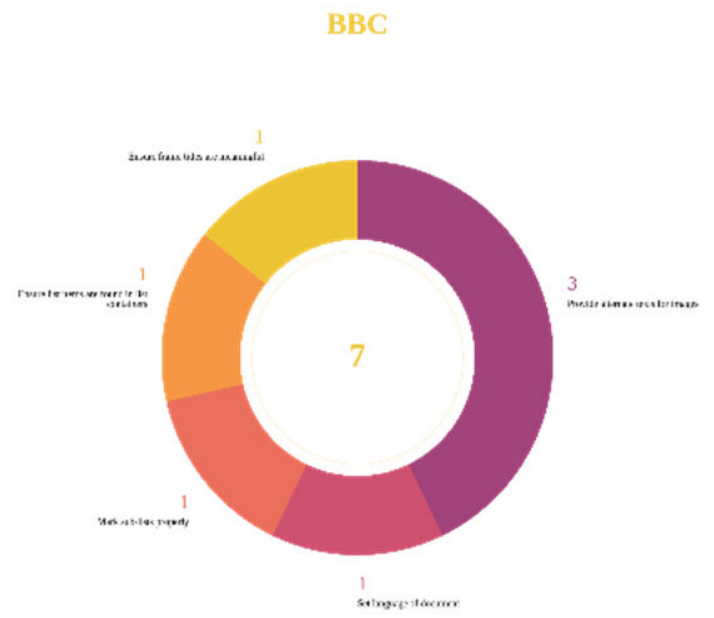 |
| **Issues Identified** | • Links open without warning<br>• Markup documents do not contain well-formed elements. | • Nonmeaningful frame titles<br>• List items not found in list containers<br>• Sub-lists not marked properly<br>• Language of document not set<br>• Alternate texts for images not provided. |
| **Alternate Suggestions** | • Avoid unnecessary creation of links that open a new window<br>• Exceptions such as "print the page" and/or "delete this" prompt should appear at minimal intervals<br>• Pages should have complete start and end tags<br>• Elements should be nested according to specification and should not contain duplicate attributes<br>• All IDs' should be made unique. | • Images within a page must be given an alternative text equivalent tag<br>• Appropriate 'alt' attribute should be added for images and poster images of videos<br>• 'alt' attribute should be concise and meaningful<br>• Logo should be marked properly<br>• Language should be set appropriately, to ensure screen-reader and other user tools to understand the same<br>• Default language for text in attribute on HTML tag should be declared. |

- ***A-Checker by Inclusive Design Research Centre*** [Release date: 19-Sept-2019]
  This tool is based on Open Accessibility Checks (OAC) providing users the ability to create their own guidelines, that can be taken care of, while checking for conformance.

| **Results** | **WHO** | **BBC** |
|---|---|---|
| **Error Chart** | 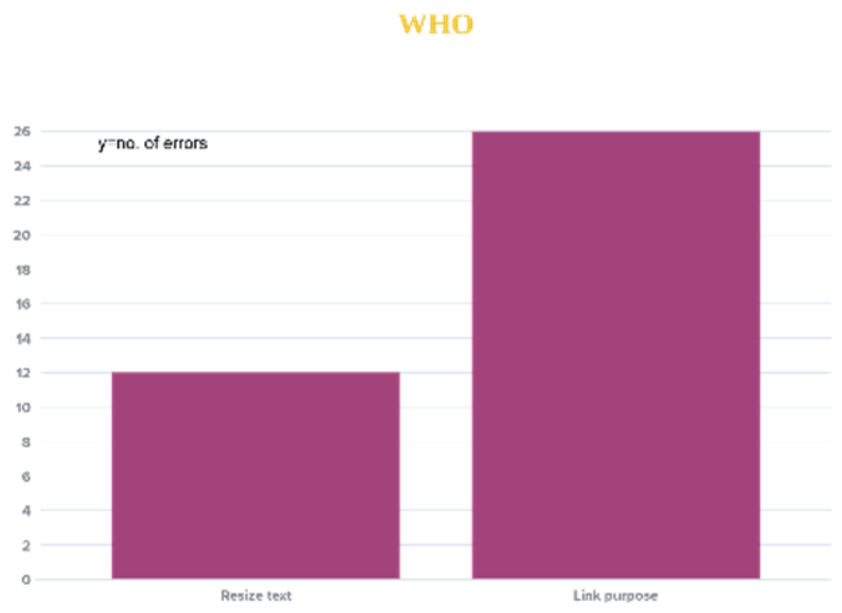 | 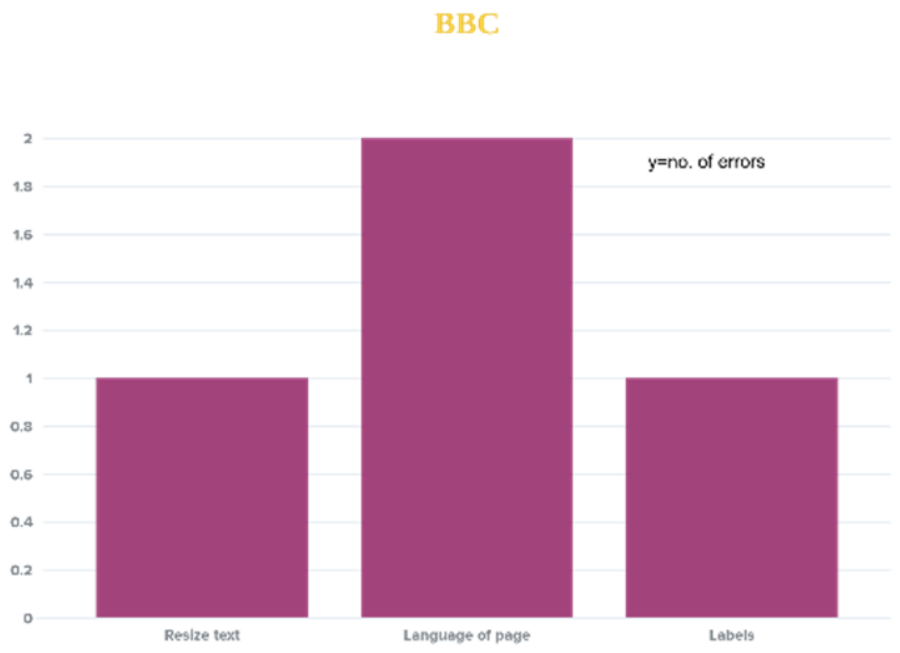 |
| **Issues Identified** | • Inappropriate size of text<br>• Links not working properly. | • Inappropriate size of text<br>• Language of page not set<br>• Labels not proper. |
| **Alternate Suggestions** | • Elements in italics should comply with guidelines by replacing with ‘em’ or ‘strong’ elements<br>• If image is used within the anchor, ‘alt’ text with a title attribute of 'a' should be added. | • Elements in italic should comply with guidelines by replacing them with ‘em’ or ‘strong’ elements<br>• For HTML documents, add ‘lang’ attribute with a valid ISO-639-1, two-letter language code to the opening HTML element<br>• For XHTML documents, add ‘lang’ and ‘xmllang’ attributes with a valid ISO-639-1 two-letter language code to the opening HTML element<br>• Input assistance should be improved<br>• Label texts should be filled appropriately<br>• Text should be added to label element to help users correct their mistake. |

- ***Functional Accessibility Evaluator 2.0 by University of Illinois at Urbana-Champaign*** [Release date: 07-Sept-2016]
  This tool evaluates webpages for WCAG 2.0 Level A & AA conformances. A unique feature of the tool, is to provide a list of rules and manual checks, required to make these websites fully accessible. It calculates the implementation score (Score= P/(P+F+MC); where P= Elements passed, F= Elements failed & MC= manual checks required) for each scanned parameter out of 100. The implementation score was found to be approximately 37 and 44 for the WHO and BBC websites respectively, which is considered below average in most cases.

| **Results** | **WHO** | **BBC** |
|---|---|---|
| **Error Chart** | 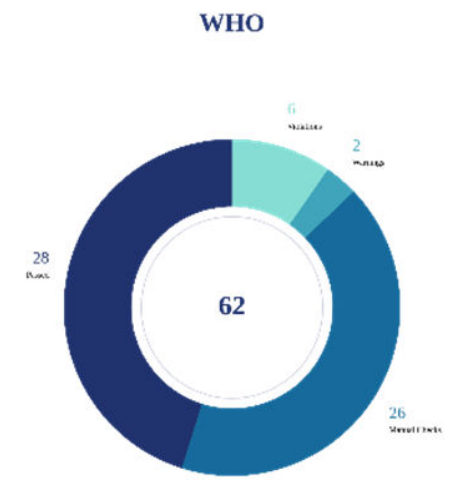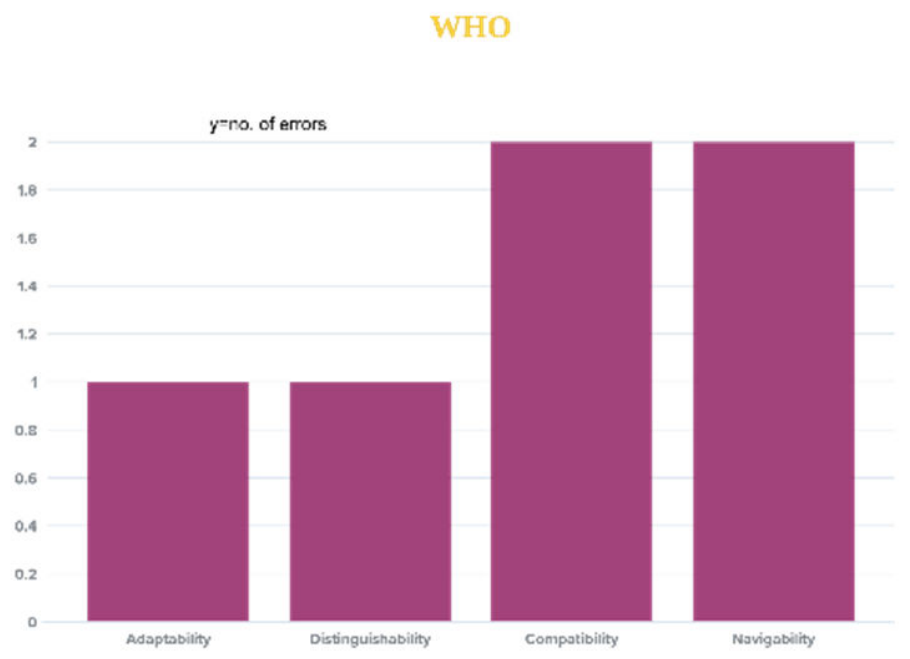 | 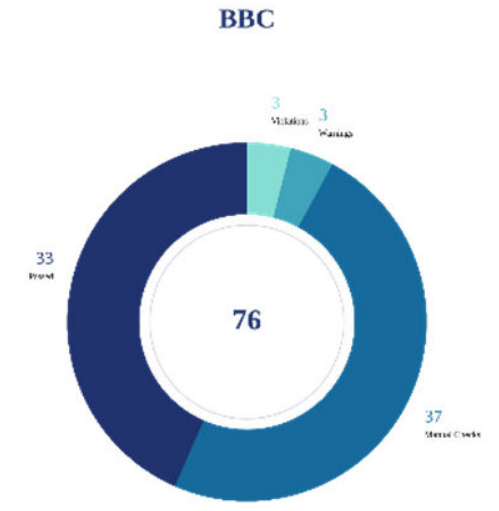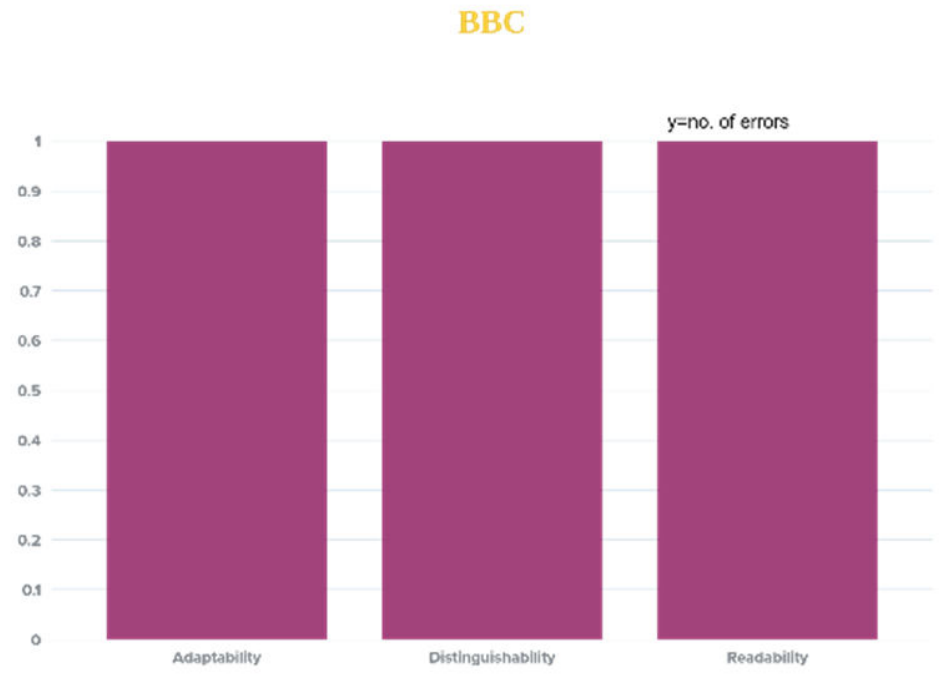 |
| **Issues Identified** | • *Not adaptable across all browsers*- parts of the webpage content are not organized and rendered in line with one another.<br>• *Not distinguishable*- parts of the webpage does not allow user to see and hear content between foreground and background<br>• *Not compatible*- the webpage does not adapt with the current and future user agents<br>• *Not navigable*- webpage fails to help users find content and determine where they are. | • *Not adaptable across all browsers*- parts of the webpage content are not organized and rendered in line with one another.<br>• *Not distinguishable*- parts of the webpage does not allow user to see and hear content between foreground and background<br>• *Not readable*- text content of the webpage is not readable and understandable. |
| **Alternate Suggestions** | • Text alternatives should be provided for any non-text content to convert into different forms such as large print, braille, speech, symbols or simpler languages<br>• Any text in images should be at least 14 points with enough contrast<br>• Need for high visibility highlighting mechanism for links and controls upon hovering of cursor<br>• Pages should be inter-linked properly with information about corresponding pages. | • Text alternatives should be provided for any non-text content to convert into different forms such as large print, braille, speech, symbols or simpler languages<br>• Content should be structured to programmatically circumscribe any assistive technology<br>• All webpages should be properly linked to each other<br>• Any text in images should be at least 14 points with enough contrast<br>• Need for high visibility highlighting mechanism for links and controls upon hovering of cursor. |

- ***Contrast checker by Acart*** [Release date: 01-Jan-2001]
  This tool evaluates a website for color contrast, based on a formula and displays checks that has passed. The color code of websites can either be provided manually or evaluated automatically by the tool.

| **Results** | **WHO** | **BBC** |
|---|---|---|
| **Error Chart** | 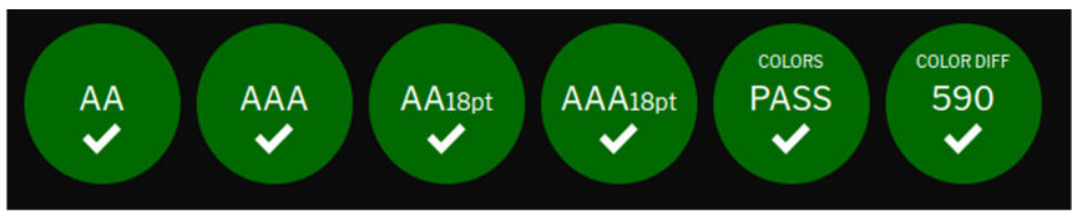<br>Foreground: 312E2D<br>Background: F4F3F3 | 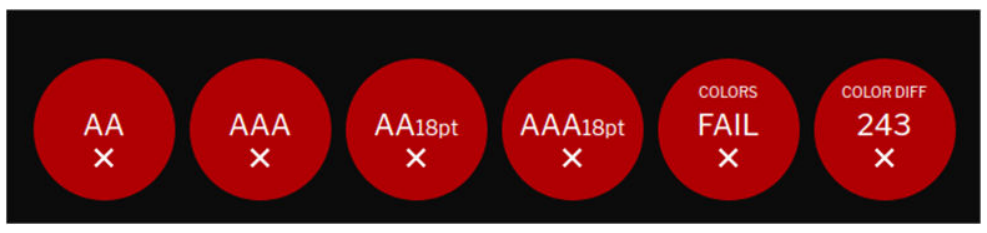<br>Foreground: C2B9B2<br>Background: 78645E |
| **Issues Identified** | None | • Does not meet W3C color contrast guidelines. |
| **Alternate Suggestions** | None | • Use 18 points (18.5px in CSS3) or 14 points bold (24px in CSS3) for text |
| | | • Contrast ratio of 3:1 should be maintained as per ISO & ANSI standards<br>• Contrast ratio of 4:5:1 should be maintained for Level AA for people with 20/40 vision, and 7:1 for people with 20/80 vision. |

- ***WAVE by WebAIM*** [Release date: 01-Jan-2014]
  This tool adds icons to a web page and marks potential accessibility concerns. Errors are marked on an interactive interface making it easy to comprehend. Red icons marked, indicate accessibility errors; yellow icons symbolize alerts; green icons symbolize accessibility features; and all light blue icons symbolize structural, semantic or navigational elements. This tool has a high CER score indicating that all errors can be identified, thereby exhibiting high performance [Alsaeedi 2020].

| **Results** | **WHO** | **BBC** |
|---|---|---|
| **Error Chart** | 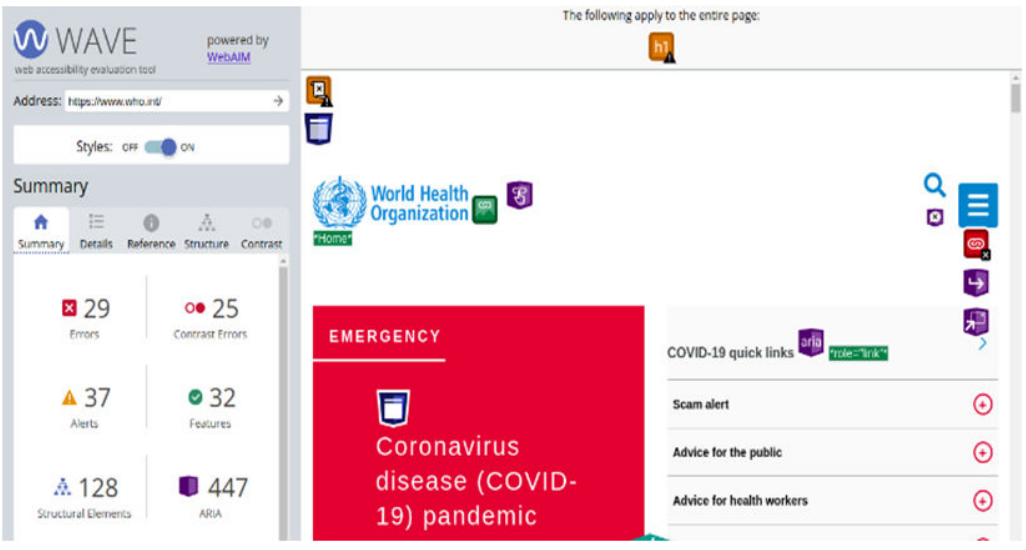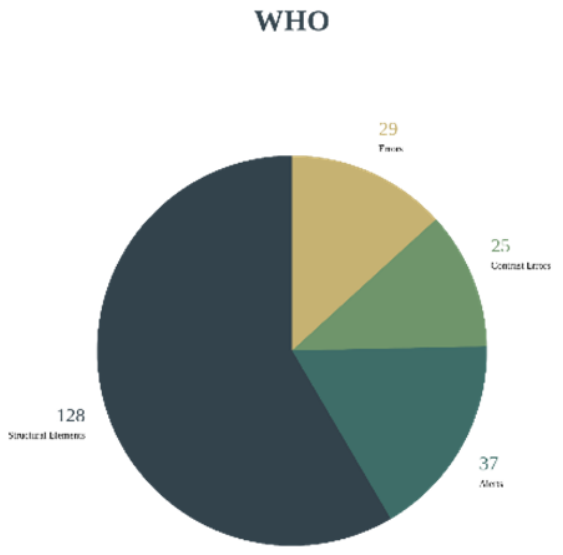 | 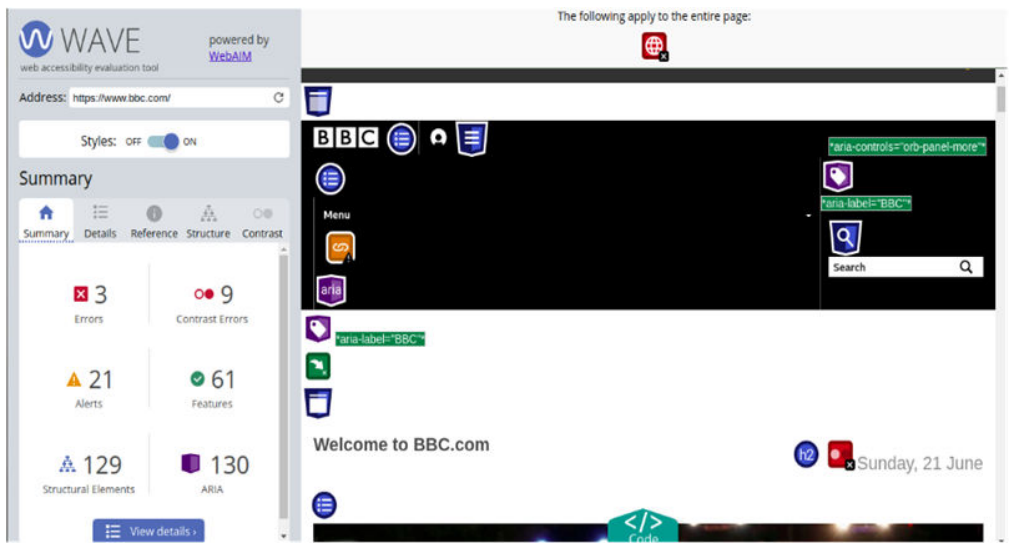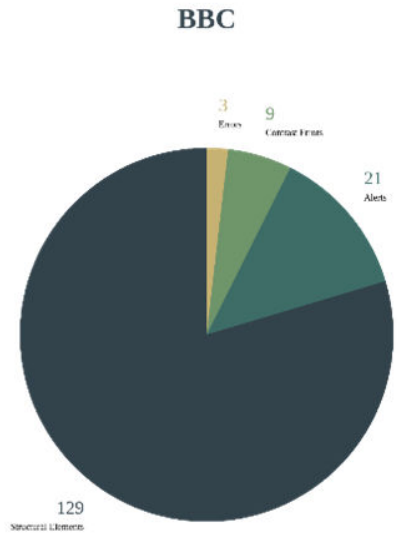 |
| **Issues Identified** | • 29 major errors<br>• 25 contrast errors<br>• 37 minor errors or alerts<br>• 128 structured elements error. | • 3 major errors<br>• 9 contrast errors<br>• 21 minor errors or alerts<br>• 129 structured elements error. |
| **Alternate Suggestions** | • For contrast, use 18 points (18.5px in CSS3) or 14 points bold (24px in CSS3) for text<br>• Contrast ratio of 3:1 should be followed as per ISO & ANSI standards<br>• Contrast ratio of 4:5:1 should be maintained for Level AA for people with 20/40 vision and 7:1 for people with 20/80 vision.<br>• HTML and CSS files should be well-structured. | |

- ***Accessibility Insights for Web by Microsoft*** [Release date: 12-Mar-2019]
  This tool checks for accessibility issues by generating a brief report of step by step guidance on improvements for a webpage.

| **Results** | **WHO** | **BBC** |
|---|---|---|
| **Error Chart** | 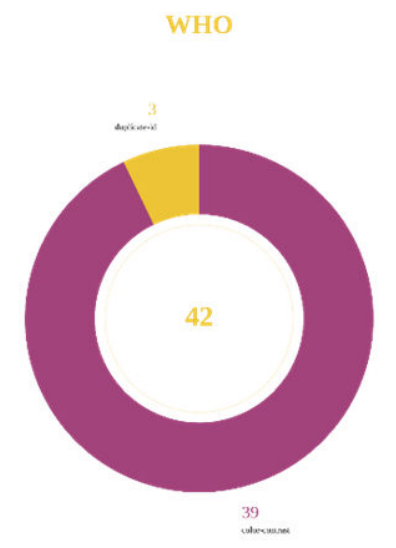 | 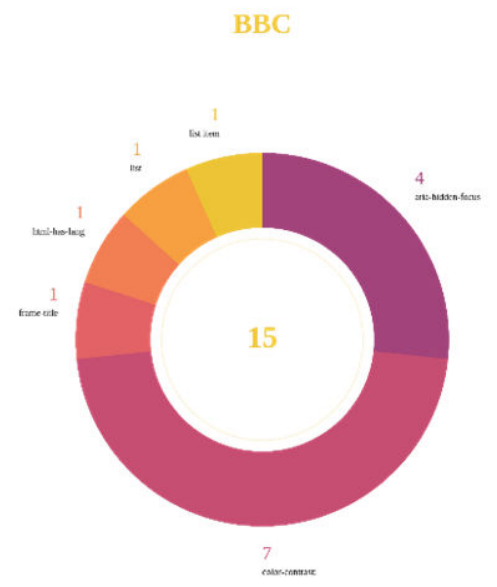 |
| **Issues Identified** | • Presence of ‘duplicate ID’ & ‘color contrast’ tags. | • Issues with ‘list item’, ‘list’, ‘html has lang’, ‘frame title’, ‘aria-hidden-focus’ and ‘color-contrast’ tags. |
| **Alternate Suggestions** | • Contrast between foreground and background color should be as per WCAG2 and Level AA guidelines<br>• Ensure unique value for each ID attribute. | • Remove all focus elements from ‘aria-hidden’ elements<br>• Contrast between foreground and background color should be as per WCAG 2 and Level AA guidelines<br>• Ensure <iframe> and <frame> elements to contain a non-title attribute<br>• Ensure each HTML document to have a lang attribute<br>• Ensure list to be structured properly<br>• List items like <li> elements should be semantically used. |

- ***Button Contrast Checker by Aditus*** [Release date: 10-Sep-2019]
  This tool checks for WCAG 2.1 compliance of all buttons and links on the webpage in just a click of a button. This tool specifically checks for contrast of buttons.

| **Results** | **WHO** | **BBC** |
|---|---|---|
| **Error Chart** | 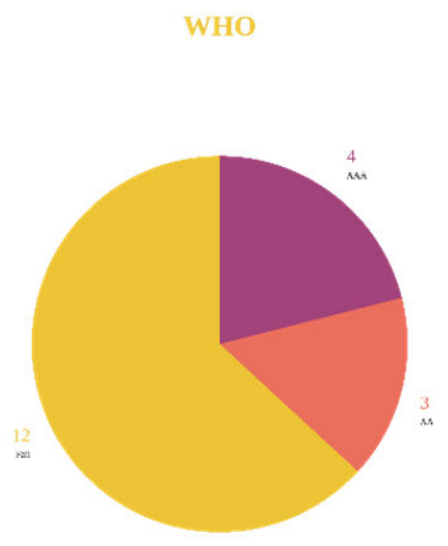 | 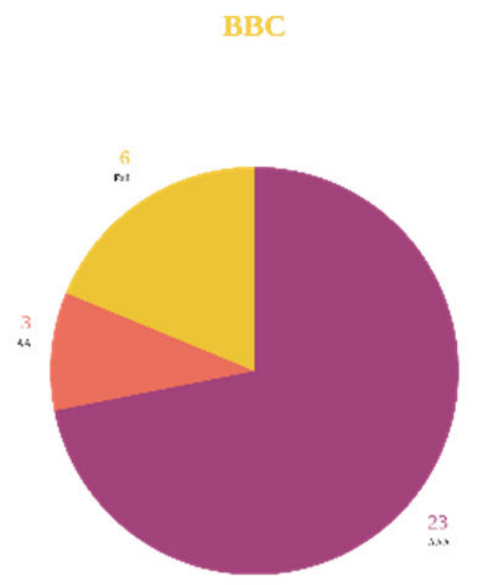 |
| **Issues Identified** | • 3 Level AA contrast guidelines are not fulfilled<br>• 4 Level AAA contrast guidelines are not fulfilled<br>• 12 contrast errors found. | • 3 Level AA contrast guidelines are not fulfilled<br>• 23 Level AAA contrast guidelines are not fulfilled<br>• 6 contrast errors found. |
| **Alternate Suggestions** | • For every button, a contrast ratio of 3:1 should be maintained as per ISO and ANSI standards.<br>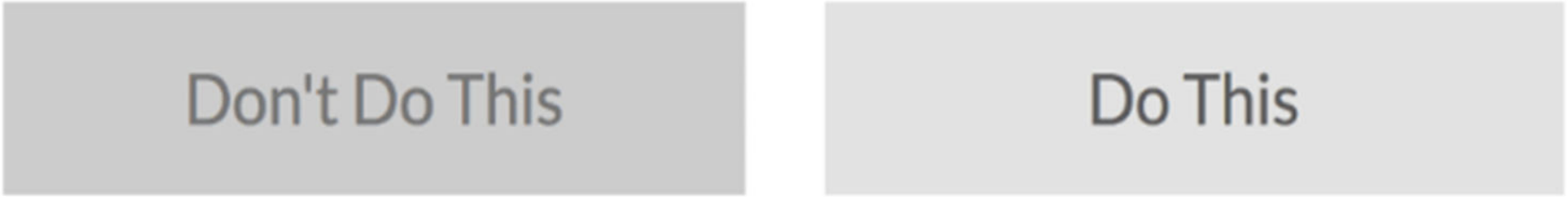<br><br>*Difference between valid and invalid color contrast* | |

- ***Siteimprove Accessibility Checker by Siteimprove*** [Release date: 14-Feb-2014]
  This tool scans individual webpages and provides a clear explanation of different issues and how to fix each issue according to the WCAG standard. It scans for restricted and password-protected pages. This tool exhibits high performance [Sulaiman et al. 2012] with a high CER score indicating that all errors can be identified precisely.

| **Results** | **WHO** | **BBC** |
|---|---|---|
| **Error Chart** | 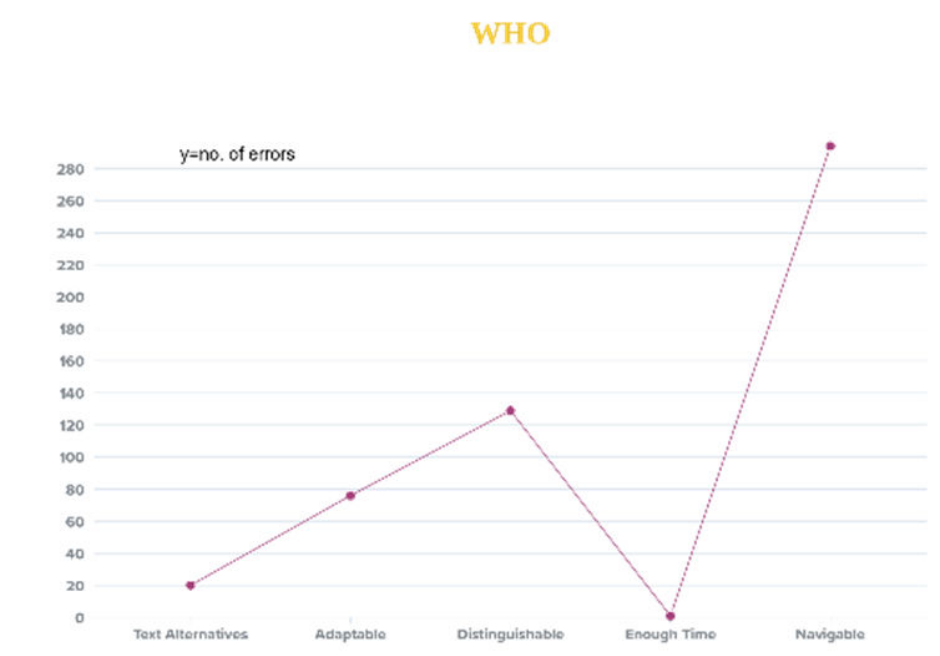<br> | 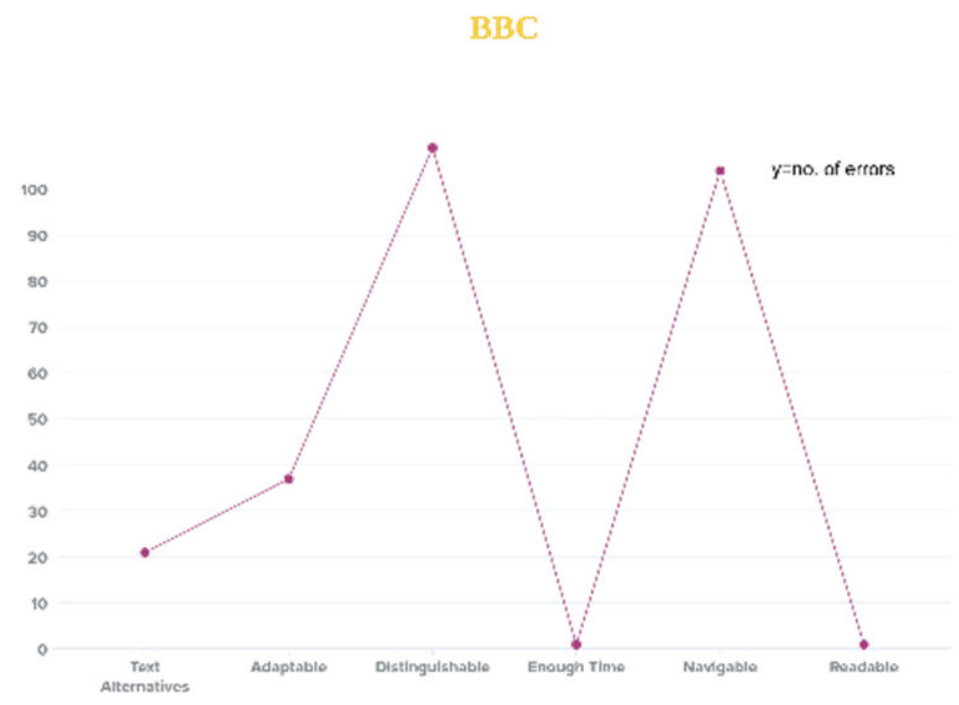<br> |
| **Issues Identified** | • Limited text alternatives are provided<br>• *Not adaptable across all browsers*- parts of the webpage content are not organized and rendered in line with one another<br>• *Not distinguishable*- parts of the webpage does not allow user to see and hear content between foreground and background<br>• Not enough time provided for user to respond to the web content<br>• *Not navigable*- webpage fails to help users find content and determine where they are. | • Limited text alternatives are provided<br>• *Not adaptable across all browsers*- parts of the webpage content are not organized and rendered in line with one another<br>• *Not distinguishable*- parts of the webpage does not allow user to see and hear content between foreground and background<br>• Not enough time provided for user to respond to the web content<br>• *Not navigable*- webpage fails to help users find content and determine where they are |
| | | • *Not readable*- text content of the webpage is not readable and understandable. |
| **Alternate Suggestions** | • Text alternatives should be provided for any non-text content<br>• Content should be structured to programmatically circumscribe any assistive technology<br>• Parts of webpage should be properly linked to one another<br>• Text in images should be at least 14 points with enough contrast<br>• Need for high visibility highlighting mechanism for links and controls upon hovering of cursor<br>• Option for adjustable timing features like pause, stop, hide, no timing, interruptions, re-authenticating and timeout should be provided<br>• Page links should have a short summary of offerings in corresponding page. | • Text alternatives should be provided for any non-text content<br>• Content should be structured to programmatically circumscribe any assistive technology<br>• Text in images should be at least 14 points with enough contrast<br>• Need for high visibility highlighting mechanism for links and controls upon hovering of cursor<br>• Option for adjustable timing features like pause, stop, hide, no timing, interruptions, re-authenticating and timeout should be provided<br>• Page links should have a short summary of offerings in corresponding pages<br>• Parts of webpage should be properly linked to one another. |

- ***Utilitia Validator by Utilitia SP. z O.O*** [Release date: 01-Jan-2011]
  This tool scans webpages and checks their accessibility. The tool checks for validity of webpage by using additional guidelines provided by the Polish Government.

| **Results** | **WHO** | **BBC** |
|---|---|---|
| **Error Chart** | 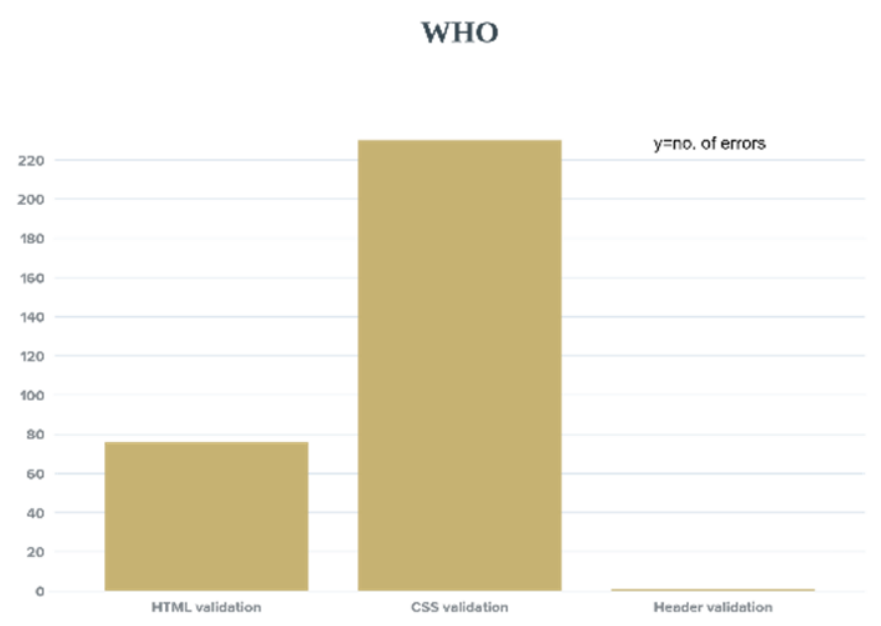  | 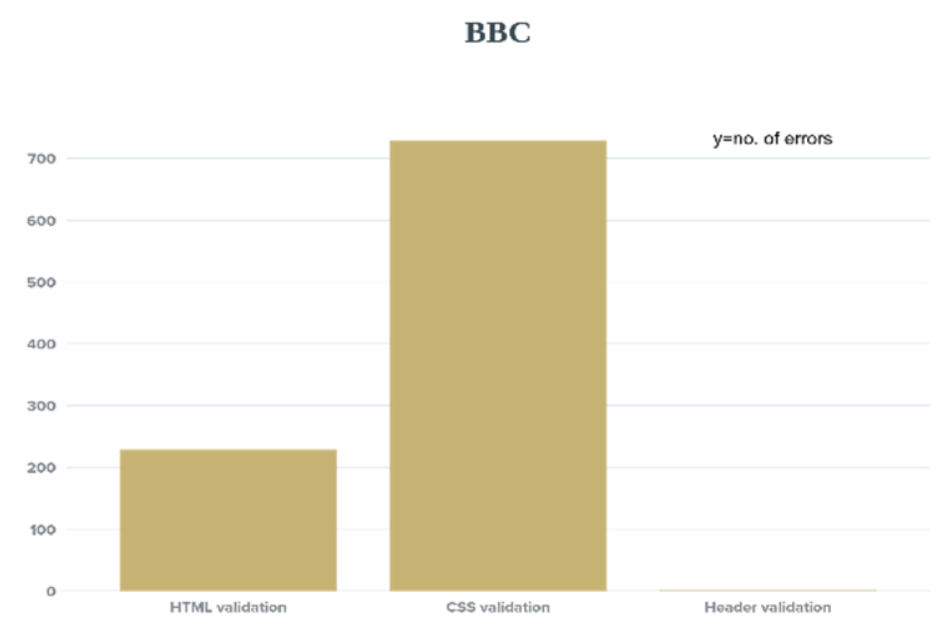  |
| **Issues Identified** | • Errors in HTML, CSS and header validation code. | • Errors in HTML and CSS code. |
| **Alternate Suggestions** | • HTML and CSS code should be properly rendered to make features available<br>• Headers should be checked and added if necessary. | • HTML and CSS code should be properly rendered to make features available. |

- ***WhoCanUse by Corey Ginnivan*** [Release date: 14-Nov-2014]
  This tool is an open source software that checks for color contrast and its effect on people with disability. This tool is specifically designed to check all 12 parameters (Table 1) for contrast correctness.

| **Results** | **WHO** | **BBC** |
|---|---|---|
| **Tool Output** | 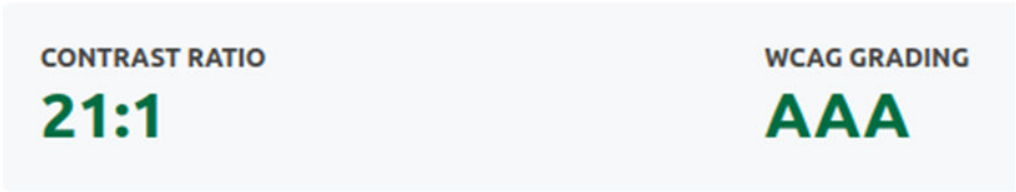  | 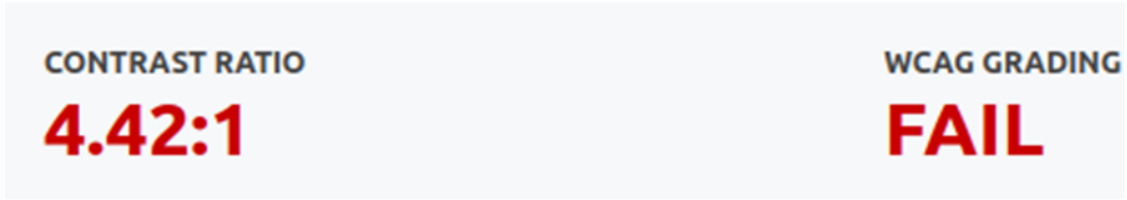  |
| **Issues Identified** | None | • Fail to meet the W3C color contrast guidelines. |
| **Alternate Suggestions** | None | • Use 18 points (18.5px in CSS3) or 14 points bold (24px in CSS3) for text<br>• Contrast ratio of 3:1 should be maintained according to the ISO & ANSI standards<br>• Contrast ratio of 4:5:1 should be maintained for Level AA as it adapts well for people with 20/40 vision and 7:1 for people with 20/80 vision. |

**Table 1: Types of Vision Parameters Checked by The Tool**

| **Vision Type** | **Population Affected** | **Definition** | **WCAG Grading** |
|---|---|---|---|
| **Regular vision (Trichromatic)** | 68% | Can distinguish all three primary colors- red, green, blue | AAA |
| **Protanomaly** | 1.3% | Trouble in distinguishing the color red | AAA |
| **Protanopia** | 1.5% | Red blind – cannot see the color red | AAA |
| **Deuteranomaly** | 5.3% | Trouble in distinguishing the color green | AAA |
| **Deuteranopia** | 1.2% | Green blind - cannot see the color green | AAA |
| **Tritanomaly** | 0.02% | Trouble in distinguishing the color blue | AAA |
| **Tritanopia** | <0.03 % | Blue blind - cannot see the color blue | AAA |
| **Achromatomaly** | <0.1% | Partial color blindness- absence of most colors | AAA |
| **Achromatopsia** | <0.1 % | Complete color blindness- can only see shades | AAA |
| **Cataracts** | 33 % | Clouding of lens in the eye affecting vision | AAA |
| **Glaucoma** | 2% | Slight vision loss | AAA |
| **Low Vision** | 31 % | Decreased and/or blurry vision (not fixable by usual means like glasses). | AAA |

After careful evaluation, we identified many issues in both websites. Websites lacked color contrast between background and foreground; lack of sign language alternatives; opening of pop-ups without proper warnings; irrelevant links; and poorly formatted markup documents and languages. By incorporating and rectifying above mentioned issues, developers can improve both websites making it accessible to maximum audience.

### 4.1.1 Summary of Results

This section presents results obtained, based on parameters considered, to evaluate the two websites using 10 web accessibility tools. The two websites were evaluated using Chrome browser (Version: 85.0.4183.83) on a Laptop (Windows 8.1, Operating System having Intel(R) Core (TM) with i5- 8250U (8th Gen) CPU @ 3.40 GHz (4 CPUs and installed memory (RAM) 8000 MB). The 10 tools are compared based on many factors, based on the WCAG guidelines. Factors such as Perceivability, Operability, Understandability, and Robustness (known as POUR factors) were considered while evaluating websites. Table 2 describes the POUR factors supported by each tool, with grey shading for rows where all four POUR criteria satisfied.

- *Perceivability:* Users can distinguish between the content in foreground and background with their senses
- *Operability:* Users can use all links, buttons, and controls that the website offers
- *Understandability:* Website should be consistent with presentation and format across all pages
- *Robustness:* Website should be compliant with the required standards. User should be able to access the website from various devices.

**Table 2: Comparison of Parameters Evaluated using POUR Factors**

| | **Parameters Checked** | | | |
|---|---|---|---|---|
| | **Perceivability** | **Operability** | **Understandability** | **Robustness** |
| **A-Tester by Evaluera Ltd** | No | Yes | Yes | Yes |
| **A-Checker by Inclusive Design Research Centre** | Yes | Yes | Yes | Yes |
| **Functional Accessibility Evaluator 2.0 by University of Illinois at Urbana-Champaign** | Yes | Yes | Yes | Yes |
| **Contrast checker by Acart** | Yes | No | No | No |
| **WAVE by WebAIM** | Yes | No | Yes | Yes |
| **Accessibility Insights for Web by Microsoft** | Yes | Yes | Yes | Yes |
| **Button Contrast Checker by Aditus** | Yes | No | No | No |
| **Siteimprove Accessibility Checker by Siteimprove** | Yes | Yes | No | No |
| **Utilitia Validator by Utilitia SP. z O.O** | Yes | Yes | Yes | Yes |
| **WhoCanUse by Corey Ginnivan** | Yes | No | No | No |

Further, tools have been compared for parameters (Table 3) like formats and browsers not supported, WCAG guidelines followed, types of automatic checks, disability checks, easy of usability and interpretation. Results for each parameter checks, are given below from Table 5 through Table 7.

**Table 3: Parameters Considered for Tool Comparison**

| Parameters Considered | Description |
|---|---|
| **Formats and Browsers Not Supported** | Check for formats and browsers that the tool cannot support. HTML, XHTML, CSS, Images, AJAX, PDF Documents; Mozilla Firefox, Opera, Apple Safari were few formats and browsers that were not supported by most WCAG tools. However, we noted that the Chrome browser was found to be supporting all the WCAG tools [Web Accessibility Evaluation Tool List 2020] |
| **WCAG Guidelines Followed** | Check if the webpage follows the WCAG 2.0 Level AA, Barrierefreie-Informationstechnik-Verordnung (BITV) guidelines |
| **Types of Automatic Checks** | Check if the tool can support a webpage which is restricted, or password protected with single or multiple webpages |
| **Disability Checks** | Check if the tool can evaluate a website for a specific target group or type of disability |
| **Ease of Usability and Interpretation** | Check for the tool to be easy, moderate or difficult in terms of usability and interpretation of results of the tool. Table 4 describes the the classification considered for ease of usability and interpretation. |

**Table 4: Criteria to Classify a Tool for Ease of Usability and Interpretation**

| Classification | Ease of Usability | Interpretation |
|---|---|---|
| **Easy** | A single link of website is provided as input to the web accessibility tool | Output provides a complete detailed report of issues and errors for a given website |
| **Moderate** | Add plugin of tool to browser followed by inserting link of website in the tool | Output provides an overview of part of webpage needed to be improved, rather than a detailed report |
| **Difficult** | Insert available script of tool into the HTML file of website. | Output does not indicate pointers to any improvement for a given issue, rather just identifies errors. |

In Table 5, the formats that are not mentioned in the formats supported section of the W3C Tool List [2020] have been considered as formats not supported, since the tool cannot evaluate the website in the said format.

**Table 5: Formats Not Supported and Browsers Supported by Tools**

| | **Formats Not Supported** |
|---|---|
| **A-Tester by Evaluera Ltd** | XHTML, CSS, Images, AJAX, PDF Documents |
| **A-Checker by Inclusive Design Research Centre** | Images, PDF Documents |
| **Functional Accessibility Evaluator 2.0 by University of Illinois at Urbana-Champaign** | XHTML, PDF Documents |
| **Contrast checker by Acart** | HTML, XHTML, CSS, AJAX, PDF Documents |
| **WAVE by WebAIM** | PDF Documents |
| **Accessibility Insights for Web by Microsoft** | XHTML, CSS, Images, AJAX, PDF Documents |
| **Button Contrast Checker by Aditus** | XHTML, CSS, Images, AJAX, PDF Documents |
| **Siteimprove Accessibility Checker by Siteimprove** | HTML, XHTML, CSS, Images, AJAX |
| **Utilitia Validator by Utilitia SP. z O.O** | Nil |
| **WhoCanUse by Corey Ginnivan** | HTML, XHTML, CSS, AJAX, PDF Documents |
| | |

**Table 6: Comparison of WCAG Criteria Covered and Automatic Checks Performed by Websites**

| | **Parameters Checked** | |
|---|---|---|
| | **Criteria Covered** | **Automatic Checks** |
| **A-Tester by Evaluera Ltd** | WCAG 2.0 Level AA | Single webpages, restricted or password protected pages |
| **A-Checker by Inclusive Design Research Centre** | WCAG 2.0, WCAG 1.0, Section 508, U.S. federal procurement standards, BITV 2.0, Italian accessibility legislation, Stanca Act, and German government standards | Single web pages, restricted or password protected pages |
| **Functional Accessibility Evaluator 2.0 by University of Illinois at Urbana-Champaign** | WCAG 2.0 Level A and AA compliance | Groups of webpages or websites |
| **Contrast checker by Acart** | WCAG 2.0 Color Contrast Success criteria | Color code manually entered |
| **WAVE by WebAIM** | WCAG 2.1, WCAG 2.0, Section 508, federal procurement standard | Single web pages, groups of web pages or web sites, restricted or password protected pages |
| **Accessibility Insights for Web by Microsoft** | WCAG 2.1, WCAG 2.0, WCAG 1.0 | Single web pages |
| **Button Contrast Checker by Aditus** | WCAG 2.1, WCAG 2.0 | Single web pages |
| **Siteimprove Accessibility Checker by Siteimprove** | WCAG 2.0, Section 508, US federal procurement standard, | Single web pages, restricted or password protected pages |
| | JIS, Japanese industry standard, Stanca Act, Italian accessibility legislation, BITV 2.0, German government standard | |
| **Utilitia Validator by Utilitia SP. z O.O** | WCAG 2.0 | Single web pages, groups of web pages or web sites |
| **WhoCanUse by Corey Ginnivan** | WCAG 2.1 | Color code manually entered |

**Table 7: Comparison of Disability Checks Covered and Ease of Usability and Interpretation of Tools**

| | **Parameters Checked** | |
|---|---|---|
| | **Disability Checks Covered** | **Ease of Usability and Interpretation** |
| **A-Tester by Evaluera Ltd** | Physical disabilities, auditory disabilities | Easy/Moderate |
| **A-Checker by Inclusive Design Research Centre** | Visual disabilities, physical disabilities | Difficult/Moderate |
| **Functional Accessibility Evaluator 2.0 by University of Illinois at Urbana-Champaign** | Visual disabilities, physical disabilities, auditory disabilities | Moderate/Easy |
| **Contrast Checker by Acart** | Visual disabilities | Moderate/Easy |
| **WAVE by WebAIM** | Visual disabilities, speech disabilities, auditory disabilities | Moderate/Difficult |
| **Accessibility Insights for Web by Microsoft** | Visual disabilities, physical disabilities | Easy/Moderate |
| **Button Contrast Checker by Aditus** | Visual disabilities | Moderate/Easy |
| **Siteimprove Accessibility Checker by Siteimprove** | Visual disabilities, speech disabilities, physical disabilities, auditory disabilities | Moderate/Moderate |
| **Utilitia Validator by Utilitia SP. z O.O** | Physical disabilities, auditory disabilities | Moderate/Difficult |
| **WhoCanUse by Corey Ginnivan** | Visual disabilities. | Moderate/Easy. |

After detailed comparison of 10 WCAG tools, we can conclude that, there is no single tool which is seamless in all aspects. However, it would be safe to say that, the **A-Checker by Inclusive Design Research Centre** is the go-to tool for those requiring a precise and elaborate review of websites on almost all of the WCAG criterions. We noted that it supports all POUR and WCAG 2.0 criteria, supports XHTML and password protected pages and the output is moderately difficult to interpret.

## 4.2 Cambridge Simulator

In addition to evaluating the WHO and BBC websites with 10 WCAG tools, we used a simulation-based approach to further identify issues and errors, that can complement the existing results. We used an inclusive performance simulator developed by Biswas et al [2012] at the Cambridge University. The Cambridge Simulator provides an overview, of how a website could be perceived by a user with different range of abilities. The simulator imitates four different types of impairments, namely: Visual, Hearing, Cognition and Motor impairments. Further details of the simulator can be found in the paper by Biswas et al. [2012]

While the web accessibility tools provided a documented report of issues in both websites, the Cambridge simulator provided a real-time perspective of webpages in the eyes of people with disabilities. Unlike the case of WCAG tools having a predefined set of parameters to evaluate a webpage, the Cambridge Simulator can take a variety of custom inputs for different types of disabilities. Thus, advantages of using the Cambridge Simulator include:

- Interface of the simulator is interactive and user-friendly
- It simulates different disabilities without requirement of real participants
- Customization of personal user profiles and parameters as input for disability check
- Multiple websites can be simulated simultaneously
- Simulation results are achieved in real time.

### 4.2.1 Results

- **Simulation for Visual Impairment**

The simulator imitates all variants of visual impairment ranging from mild vision acuity loss, severe visual acuity loss, red-green color blindness, glaucoma, macular degeneration and diabetic retinopathy [Rubin et al. 2001, JE 1984]. Results from the simulator (Figure 2) noted that, both websites possessed lack of quality of images with poor color contrast, making the websites difficult to use by mild visual impairment.

- **Simulation for Hearing Impairment**

People with visual impairment, may find it optimal to hear the content displayed on a webpage, to help understand the website. The Cambridge simulator simulates hearing impairment with conductive and sensory hearing, ranging from mild to severe (Figure 3). Results found a noticeable difference in the output audio file, for podcast and speech, as perceived by the listener. For both websites, the audio was barely coherent when deficit of the disease was set to ‘severe’. However, when deficit was set to ‘mild’, the audio was coherent to a certain extent. Additionally, we found that the audio track of the WHO website was not processed before it was put on the website, leading to unclear sound, while the BBC website was comparatively better when compared to the WHO website. The resulting audio files are given in- https://drive.google.com/drive/folders/1OEjLKSwvqZ9UPtrvaWO7kc_nxALHKksR

- **Simulation for Mobility Impairment**

We ran the mobility impairment simulator for a profile with mild Parkinson’s Disease that results in hyperkinetic motor impairment. Most links and news items in both websites found to have adequate inter-element spacing except the pus (+) icons on WHO webpage found too small to select by person with motor impairment with predicted movement time of more than 3 secs (Figure 2).

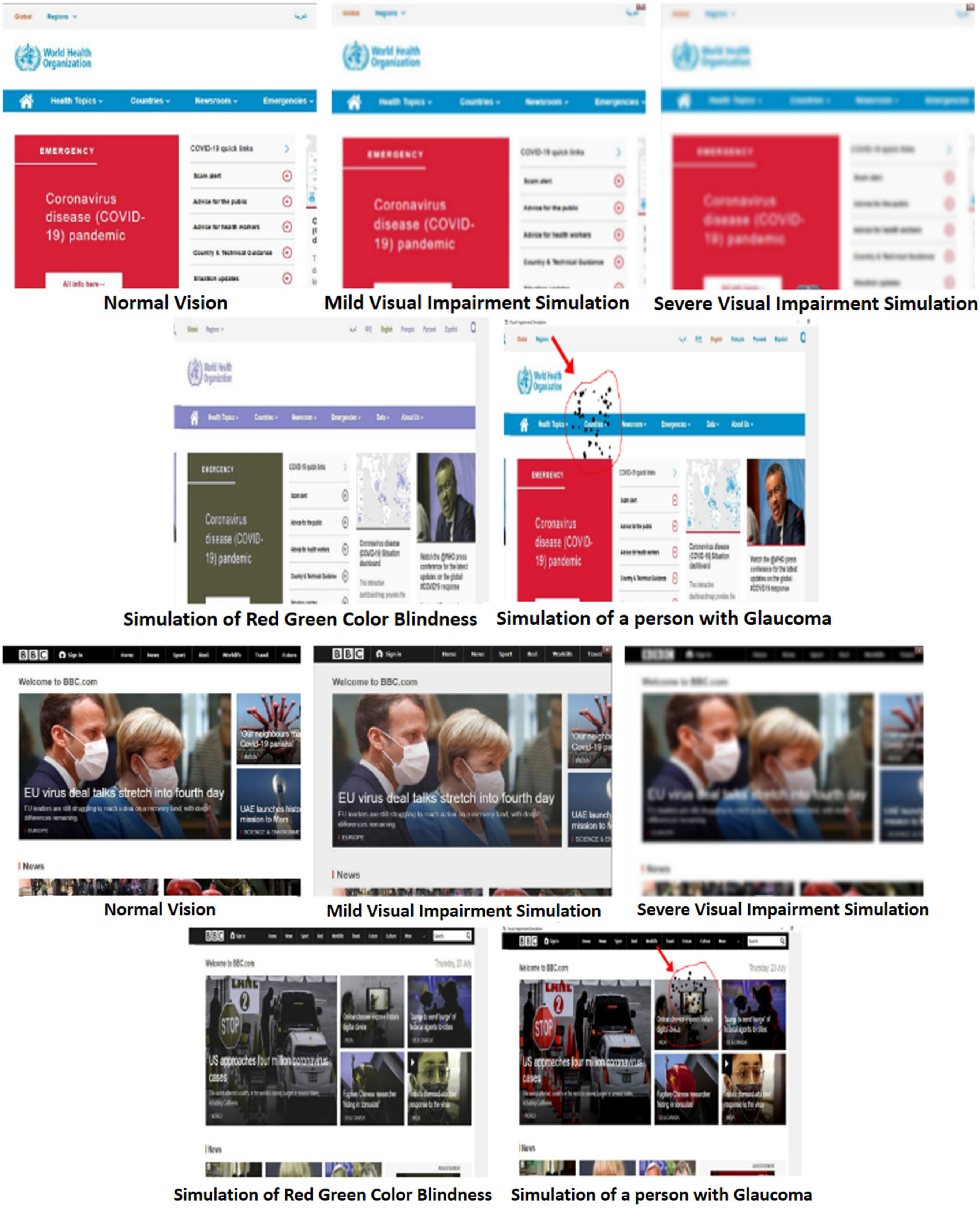

**Figure 1. Output for WHO and BBC Websites**

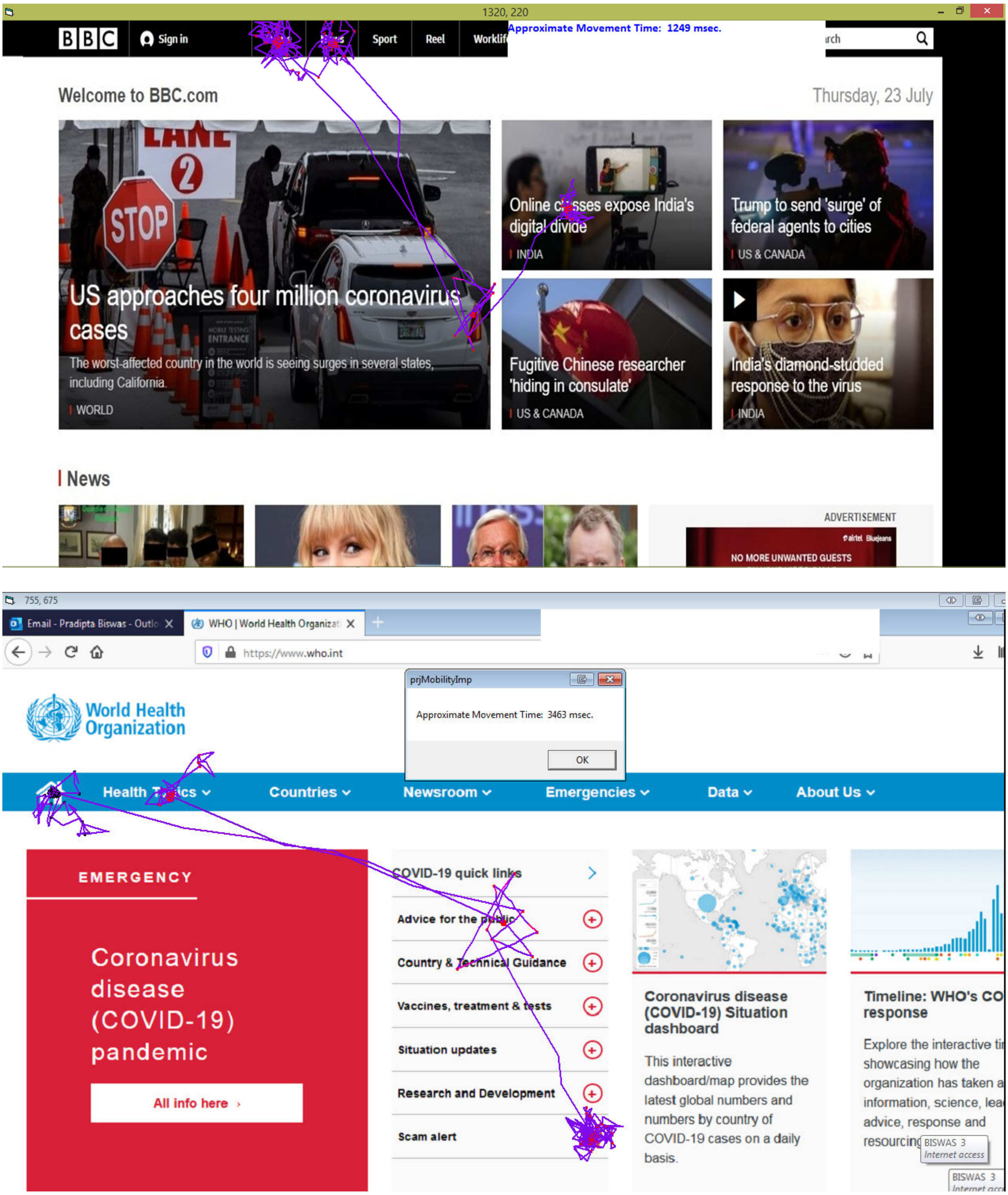

**Figure 2. Output from Motor Impairment Simulator for WHO and BBC and WHO Websites**

## 5 Overall Discussion

This paper initially used 10 web accessibility tools to evaluate landing pages of two important websites- the WHO and the BBC. We compared these 10 tools based on POUR factors for website evaluation. Evaluation of both websites revealed many scopes of improvements. In the WHO website, links open without any warning; markup document do not contain well-formed elements; size of text do not fulfill the required criteria; presence of duplicate ID in mark-up languages; header violation in HTML and CSS code were found noticeable. In the BBC website, most frame titles were not meaningful; list containers for list items were missing; sub-lists were not marked properly; language of the document was not set; alternate texts for images were not provided; size of text and labeling did not meet the WCAG criteria; specified color contrast guidelines were not followed; there was violation with respect to ‘Html-has lang', 'frame title' and 'aria-hidden focus'; insufficient time to read the content and so on.

Compared to a study by Brajnik et al. [2011], we used automated tools instead of manual evaluation with human experts, as it provided accurate and fast results with respect to syntactic issues in the markup language of a website. A study by Alsaeedi [2020] considered only two tools leading to incomplete evaluation of webpage. In our study, we used 10 different WCAG tools for website accessibility evaluation, thereby making our results more reliable. Additionally, we used a simulator [Biswas et al. 2012] simulating interaction of a wider range of abilities of users compared to existing studies [Devi 2011, Kumar and Owston 2016].

## 6 Common User Profile Format

The WHO and BBC landing pages are examples of well-designed and accessible websites. However, we may note that it is often difficult to cover requirements of a wide range of abilities of users. A one-size-fits-all approach is difficult to implement while developing different versions of the same website for people with different range of abilities and is often not scalable for large websites. Researchers already explored ways to adapt the same website differently for different users based on a user profile. The SUPPLE project [Gajos 2007] at University of Washington, Inclusive User Model [Biswas 2012] described in previous section in the context of simulation based evaluation, IBM Web Adaptation technology [2020] and AVANTI browser [Stephanidis 2003] are notable examples, mostly working for people with different range of visual and motor impairment. A user profile is an essential component for any personalization or adaptation.

From 2010 onwards, there were various attempts among researchers to create a common user profile format. The EU VUMS (Virtual User Modelling and Simulation) cluster [2020] took an ambitious attempt to publish an exhaustive set of anthropometric, visual, cognitive, auditory, motor and user interface related parameters for adapting man-machine interfaces of automobile, consumer electronics, audio-visual media and so on. A technical report from ITU Focus Group on Smart TV [2020] published a much compact set of parameters for creating user profile for smart TV.

As computers turn more ubiquitous with advancement of electronic technology, presently users including people with disabilities access information through multiple devices each having different set of applications and software platform. Ideally, an accessibility service should be provided to all devices and applications irrespective of underlying hardware. Responsive design of website can be considered as an example of automatic adaptation of layout based on screen size and platform of deployment. However, information about users is essential to personalize content and layout of an audio-visual media with respect to range of abilities of users. A user profile can be defined as an instantiation of a user model while a user model can be defined as a machine-readable description of user. In the context of personalization and accessibility, a common user profile format may have the following advantages:

1. Personalizing user interface content and layout for different applications after creation of a single user profile
2. Offering accessibility services to all devices and platforms after creation of a single user profile
3. Sharing personalized content and interface across different platform and devices to improve usability
4. Adapting quality of accessibility services (e.g. font size of subtitle) across multiple media
5. Sharing personalization metadata among service providers like website or content developers.

Figure 3 below shows examples of interface adaptation across multiple devices and platforms using a common user profile format. It may be noted in the picture that color contrast, font size, inter-element spacing of icons is adjusted across smart TV, desktop and laptop computers, smartphones and low-end mobile phones based on a common user profile.

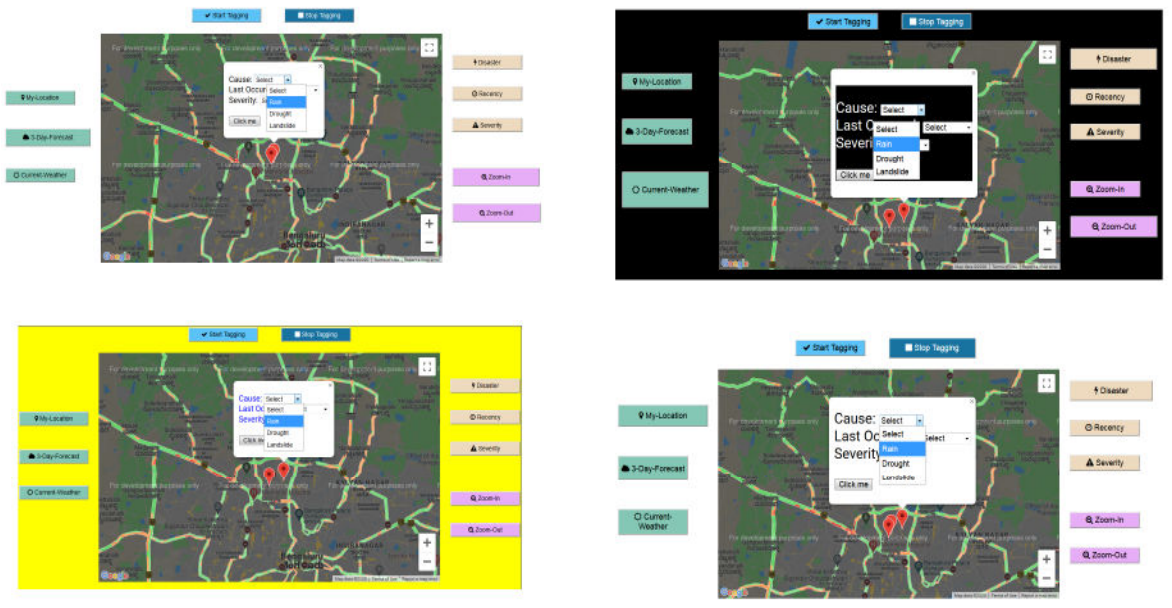

Font size and colour contrast adaptation for disaster warning application in desktop / laptop computer

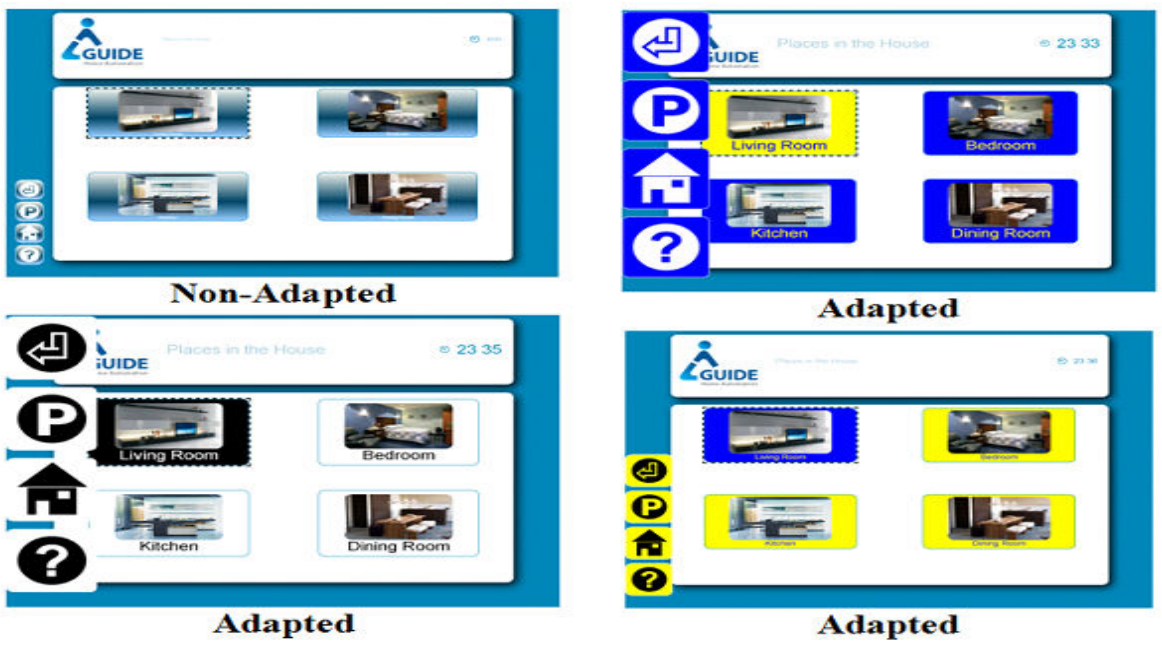

Home automation application for smart TV

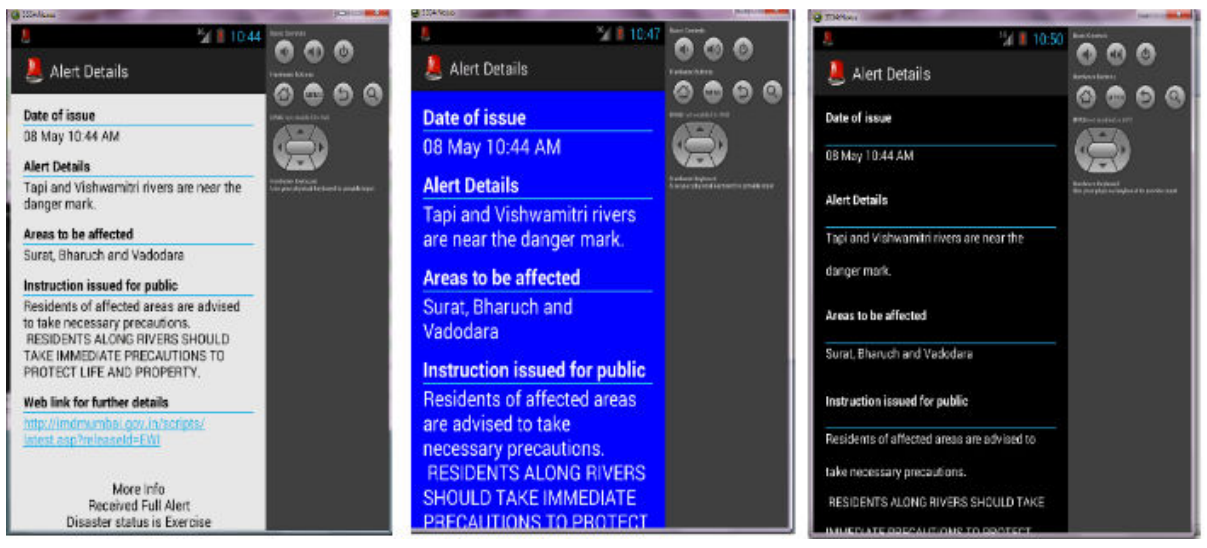

Font size and colour contrast adaptation for disaster warning application in smartphone

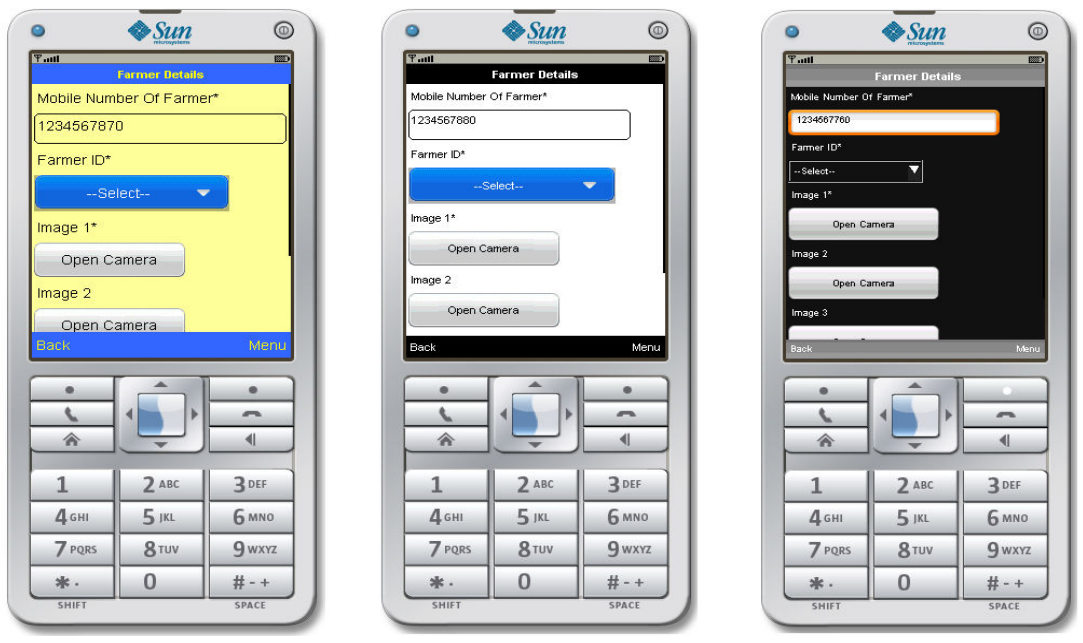

e-Agri application for low end mobile phone

**Figure 3. Example of Interface Personalization Using Common User Profile Format**

However, sharing information about user always possess security risk and unintended use not authorized by end users. Implementation of the common user profile should take care of security aspect and local regulation and legislation. If the format and details of common user profile is agreed and shared among service providers, then sharing of actual content may not be necessary, rather the personalization algorithms can run on user profile stored on local machines. Standardization and sharing of only the format definition across service providers will enable personalization without taking the risk of sharing details of individual user.

## 7 Conclusion

This paper analyzed a set of 10 WCAG evaluation tools on two popular webpages and compared the output from these tools. The webpages were evaluated to check whether they followed the WCAG guidelines by meeting all three conformance levels. While the automatic accessibility tools are easy to apply, it may be noted that their output are not equally comprehensible. There are many differences among the syntax of checking and there is not a single winning candidate among tools. Certain aspects of usability could not be captured by mere syntax checking of the tools and required thorough analysis. In this context, we presented a simulation-based approach and showed how it can complement existing WCAG tools. Finally, we presented a concept of common user profile format that can personalize websites across devices and platforms based on a common understanding of users' range of abilities.